\documentclass[aps,prl,twocolumn,groupedaddress,showpacs,showkeys,superscriptaddress,nofootinbib]{revtex4-1}%
\usepackage{graphics}%
\usepackage{amsmath}
\usepackage{mathrsfs} %script fonts
\begin{document}%
%----------------------------------------------------------------
% Definitions needed for the heading
%-------------------------------------------------------------------------
\def\Barcelo{Barcel\'o}
%-------------------------------------------------------------------------
\title{Minimal conditions for the existence of a Hawking-like flux}
%-------------------------------------------------------------------------
\author{Carlos \Barcelo}
\affiliation{Instituto de Astrof\'{\i}sica de Andaluc\'{\i}a, IAA--CSIC,
Glorieta de la Astronom\'\i{}a, 18008 Granada, Spain}
%-------------------------------------------------------------------------
%-------------------------------------------------------------------------
\author{Stefano Liberati}
\affiliation{SISSA/International School for Advanced Studies, Via Bonomea 265,
34136 Trieste, Italy \emph{and} INFN, Sezione di Trieste}
%-------------------------------------------------------------------------
%-------------------------------------------------------------------------
\author{Sebastiano Sonego}
\affiliation{Universit\`a di Udine, Dipartimento di Fisica, Via delle Scienze 208, 33100 Udine, Italy}
%-------------------------------------------------------------------------
%-------------------------------------------------------------------------
\author{Matt Visser}
\affiliation{School of Mathematics, Statistics, and Operations Research,
Victoria University of Wellington, New Zealand}
%-------------------------------------------------------------------------
\date{25 November 2010;
\LaTeX-ed \today}
%-------------------------------------------------------------------------
\bigskip
%-------------------------------------------------------------------------
\begin{abstract}
%-------------------------------------------------------------------------

We investigate the minimal conditions that an asymptotically flat general relativistic spacetime must satisfy in order for a Hawking-like Planckian flux of particles to arrive at future null infinity.  We demonstrate that there is  no requirement that any sort of horizon form anywhere in the spacetime.
We find that the irreducible core requirement is encoded in an approximately exponential ``peeling'' relationship between affine coordinates on past and future null infinity. 
As long as a suitable adiabaticity condition holds,  
then a Planck-distributed Hawking-like flux will arrive at future null infinity with temperature determined by the $e$-folding properties of the outgoing null geodesics.
The temperature of the Hawking-like flux can slowly evolve as a function of time. We also show that the notion of ``peeling" of null geodesics is distinct,
and in general different, from the usual notion of ``inaffinity" used in Hawking's definition of surface gravity.

%-------------------------------------------------------------------------
\end{abstract}
%-------------------------------------------------------------------------
%-------------------------------------------------------------------------
%-------------------------------------------------------------------------
\maketitle
%-------------------------------------------------------------------------

%----------------------------------------------------------------
% Local defines
%----------------------------------------------------------------
\newcommand{\scri}{\mathscr{I}}
\newcommand{\sun}{\ensuremath{\odot}}%
%----------------------------------------------------------------
\def\e{{\mathrm e}}%
\def\g{{\mbox{\sl g}}}%
\def\Box{\nabla^2}%
\def\d{{\mathrm d}}%
\def\R{{\rm I\!R}}%
%--------------------------------------------------
\def\ie{{\em i.e.\/}}%
\def\eg{{\em e.g.\/}}%
\def\etc{{\em etc.\/}}%
\def\etal{{\em et al.\/}}%
%--------------------------------------------------
%----------------------------------------------------------------
%----------------------------------------------------------------

Ever since Hawking's 1974 discovery of the existence of a quantum-physics-induced steady Planckian flux of particles in black hole spacetimes~\cite{hawking1, hawking2}, there has been a steady stream of papers that seek to re-derive this effect in many different ways --- typically with a view to understanding what parts of the usual derivation are truly essential, and what parts can be dispensed with, or as overall consistency checks on the entire formalism~\cite{unruh, tipler, israel, bardeen, york, hajicek, grove, entropy, essential}. In the strict general relativistic scenario adopted in this paper (no modified dispersion relations), Hawking radiation is derived assuming (i) that an event horizon forms, and (ii) that the subsequent exterior geometry is static.  However, one may be interested in either considering quasi-black holes, where (i) fails~\cite{fate, small-dark, revisit, ashtekar-bojowald, disinformation, hayward}, or following the evolution of a black hole during evaporation, where (ii) fails.  We present a formalism where one can deal with both these cases.

 One particularly important result, due to Hajicek~\cite{hajicek}, is that the existence of a strict event horizon is \emph{not} necessary, and that a long-lived apparent horizon is quite sufficient to generate the Hawking flux (see also~\cite{grove, entropy, essential}).  More recently, the present authors have developed some ``analogue spacetimes''~\cite{unexpected, acoustic, LRR} for which a Hawking flux is generated even in the absence of a trapping/apparent horizon~\cite{trapping, trapping2}, which has prompted us to undertake a thorough reassessment of the situation.  Inspired in particular by the work of Hu~\cite{hu}, we focus on the existence of an (in our case, approximate) exponential relation between the affine parameters on past and future null infinities as the necessary and sufficient condition for generating a Hawking flux. 

%\bigskip
%\enlargethispage{20pt}

\noindent\underline{\emph{Structure of null infinity}:} 
Consider an asymptotically flat spherically symmetric spacetime with a Minkowskian structure in the asymptotic past. (The discussion that follows applies equally well to any number of spatial dimensions and can easily be generalized to deal with acoustic spacetimes in 1+1 dimensions having two asymptotic regions~\cite{njp}.) In the $\{t,r\}$ sector of the geometry we  
define an affine parameter $W$ on $\scri^-$, and use it to label the null curves travelling towards the centre of the body. Similarly, $u$ is taken to be an affine parameter on $\scri^+$, used to label the null rays  travelling away from the central body.  The independent coordinates $\{W,u\}$ provide a double-null cover of the relevant parts of spacetime (the domain of outer communication).

As is standard, one can define a canonical functional relationship connecting $\scri^-$ with $\scri^+$ by using null curves that reflect off the centre at $r=0$. This relation can be expressed as
\begin{equation}
U = p(u);  \qquad u = p^{-1}(U), 
\end{equation}
where the labels $\{U,u\}$ are now no longer to be thought of as independent coordinates but, since we have explicitly linked them via the function $p(\cdot)$, as different ways of labelling the same null curve once it is reflected through the origin.
It is to be understood that $p^{-1}(\cdot)$
need not be defined on all of $\scri^-$ if a true event horizon indeed forms; however this function will certainly be well defined on those parts of $\scri^-$ that lie in the domain of outer communication.
We shall soon see that the function $p(\cdot)$, or equivalently its inverse,
is sufficient to encode all the relevant physics of Hawking radiation. Specifically, let us choose a reference null curve completely traversing the body. It is labelled by $u_*$ on its way out of the body, and by $U_*$ on its way in.  We want to use ``local'' information from the vicinity of this reference null curve to study Hawking-like radiation that reaches $\scri^+$ in the vicinity of $u_*$.

%\bigskip

\noindent\underline{\emph{Exponential representation}:} 
Let us \emph{define}
\begin{equation}
\kappa(u) \equiv  
-{\ddot p(u)/ \dot p(u)}.
\end{equation}
Around the null curve labelled by $u_*$ this definition can be integrated to yield the exact result
\begin{equation}
U = U_* 
+ C_* \;  \int^u_{u_*}  
\exp\left[  - \int^{\bar u}_{u_*} \kappa(\tilde u) \; d\tilde u \right] d \bar u,
\end{equation}
for some constant $C_*$.  At this stage this is just an alternative way of writing the function $p(u)$ in terms of another function $\kappa(u)$. Note that this formalism continues to make perfectly good sense for $\kappa\to 0$ where it implies a linear relation between $u$ and $U$.
Now in any  sufficiently small interval around $u_*$ one can always approximate $U=p(u)$ by 
\begin{eqnarray}
U &\approx&
U_* 
+ C_* \;  \int^{u}_{u_*} 
 \exp\left[  -\kappa_* \; \{\bar u-u_*\}  \right] 
d\bar u,
\end{eqnarray}
where $\kappa_*=\kappa(u_*)$. Integrating once again,
\begin{eqnarray}
U 
&\approx&
U_* 
- {C_*\over\kappa_*}  \;  \left\{ 
 \exp\left[  -\kappa_* \; \{u-u_*\}  \right] 
- 1 \right\}
\nonumber
\\
&=&
U_H^*
- A_* \;  \exp\left[  - \kappa_* u \right],
\label{E:exp}
\end{eqnarray}
where we define constants
\begin{equation}
U_H^* = U_* + {C_*\over \kappa_*}
\qquad \hbox{and} \qquad
A_* =  {C_*\over\kappa_*} \; e^{\kappa_*\,u_*}.
\end{equation}
One can also invert this ``exponential approximation''  to give the perhaps more common ``logarithmic approximation''
\begin{equation}
u \approx  - {1\over\kappa_*}\;\ln\left\{\; {U_H^*-U\over A_*}\;\right\},
\end{equation}
which is valid in the proximity of $U_*$.

The essence of this approximation is to replace $\kappa(u)$ by $\kappa_*$. This can always be done in the region defined by 
\begin{equation}
\left|\int_{u_*}^u \left[ \kappa(\bar u)-\kappa_* \right]\,d \bar u\right|\leq \epsilon^2 \ll 1.
\end{equation}
Under suitable technical assumptions this can be replaced by
\begin{equation}
|\dot \kappa(u_*) |\; (u-u_*)^2\leq \epsilon^2 \ll 1,
\end{equation}
(for mathematical and physical details see~\cite{details}).
Defining $ \dot \kappa_*=  \dot \kappa(u_*)$, the region in which the exponential approximation $\kappa(u)\approx\kappa(u_*)$
is valid can be equivalently expressed as 
\begin{equation}
\label{E:xx}
|u-u_*| \leq {\epsilon |\dot\kappa_*|^{-1/2}} \ll |\dot\kappa_*|^{-1/2}.
\end{equation}
The relevant interval 
\begin{equation}
\label{E:Sp}
{\cal S}_+ = \left(u_* - {\epsilon |\dot\kappa_*|^{-1/2}}, \, u_* + {\epsilon |\dot\kappa_*|^{-1/2}} \right),
\end{equation}
has  a counterpart  ${\cal S}_- = p({\cal S}_+)$: 
\begin{equation}
{\cal S}_- = \left(U_H^* - {C_*\over\kappa_*}\; e^{\epsilon\kappa_* |\dot\kappa_*|^{-1/2}}
\,,
U_H^* - {C_*\over\kappa_*}\; e^{-\epsilon\kappa_* |\dot\kappa_*|^{-1/2}}
\; \right),
\label{U-int}
\end{equation}
which defines what we mean by ``in the proximity of $U_*$''.

%\bigskip

%\enlargethispage{20pt}

\noindent{\underline{\em Adiabatic approximation and the Hawking flux}\/:} 
Since we know the exponential approximation is valid for $u\in {\cal S}_+$, consider some wave packet that arrives at $\scri^+$ with compact support in ${\cal S}_+$.  Such a wave packet cannot tell the difference between the (unknown) exact relation $U = p(u)$ and the ``exponential approximation" defined above. Working with such  wave packets
one can attempt to apply the standard Hawking calculation~\cite{hawking1, hawking2, unruh} to derive a time-dependent Bogoliubov coefficient $\beta(\omega, u_*; \kappa_*)$ relevant to this particular time interval. 
At a first glance it seems that this calculation should give us a Planckian spectrum with a time-dependent Hawking temperature 
\begin{equation}
k_B \; T_H(u_*) = \hbar \; {\kappa(u_*)\over2\pi}.
\end{equation}
However, this is not quite true. To obtain this result one needs to satisfy one further important condition~\cite{details}: The width of the wave packets used, $\Delta \omega$, has to be much smaller than the frequencies at the peak of the Planck spectrum, \ie,  $\Delta \omega \ll \kappa_*$.
But there is always an inverse relation between the temporal  and frequency resolutions of a wave packet $\Delta\omega\gtrsim1/\Delta u$, and Eq.~(\ref{E:Sp}) implies $\Delta u \sim \epsilon  |\dot\kappa_*|^{-1/2}$. Thus, to recover a Planckian emission at $u_*$, the following \emph{adiabatic condition} has to be satisfied:
\begin{equation}
{|\dot \kappa_*| \over \kappa_*^2} \ll \epsilon^2 \ll 1.
\end{equation} 
Only some particular functions $p(u)$ satisfy this condition, and typically only in specific $u$
regions. We shall subsequently verify that for realistic black holes this region is ``large enough''.

Once we are sure that the adiabatic condition is in place, Hawking's (at first glance seemingly absurd) extrapolation of his version of the  exponential approximation to ``all time", (equivalent to Eq.~(\ref{E:exp}) above), can simply be inserted into any one of the usual derivations of the Hawking effect, and (subject to the qualifications above) the standard result follows~\cite{hawking1, hawking2, unruh}. We have explicitly checked the details of this claim by calculating the Bogoliubov coefficients in four independent ways: By using the Klein--Gordon inner products on both $\scri^-$ and $\scri^+$, and by evaluating the relevant integrals either in terms of Gamma functions, or by using a variant of the stationary phase technique~\cite{details}. The final result is a time-dependent but slowly varying Hawking temperature.

Once in an adiabatic regime some observations are in order. The $U_H^*$ in Eq.~(\ref{E:exp}) is \emph{not} in general the location of the horizon. It is instead the best estimate (based on what you can see locally at $u_*$) of where a horizon \emph{might be likely to form} if the relation between $U$ and $u$ keeps $e$-folding in the way it is at $u_*$. There is no actual implication that a strict horizon (or indeed any sort of horizon) ever forms, only that it ``looks like'' a horizon might form in the not too distant future. If a strict stationary event horizon forms, the exponential approximation with a fixed asymptotic value for $\kappa_*$, namely $\kappa_*=\kappa_H$, would be valid from some $u_0$ up to arbitrarily large values of $u$. Then, $U_H$ would signal the location of the event horizon. However, in strictly geometric terms (without considering Einstein equations or other dynamical conditions), the adiabatic approximation could be perfectly valid in evaporating configurations even if no horizon of any type ever forms. 

%\bigskip

\noindent{\underline{\em Normalization}:} 
To fix the parameters in the exponential approximation of Eq.~(\ref{E:exp}) we have had to fix the overall normalization of $u$ and $U$. This is done in two steps. First we demand  
$p(u\to-\infty) \to u$,
to make sure that in the infinite past (\ie, before collapse) $\scri^-$ and $\scri^+$ are connected in a simple sensible way: $U = u$.
That is,
${d U/ du} = \dot p(u) \to 1$  \hbox{as} $u \to -\infty$.
Next, given the assumed asymptotic flatness, we pick an overall scale to set $u=t-r$ in terms of the asymptotic time and space coordinates.

A consequence of this normalization condition is that
\begin{equation}
C_* =
\exp\left[  -\int_{-\infty}^{u_*} \kappa(\tilde u) \; d\tilde u \right],
\end{equation}
so that $C_*$ depends on the entire past history of $\scri^+$ (the history of the collapse in Hawking's language).  In fact, noting that $\delta U = C_* \; \delta u$, it is easy to see that $C_*$ is the Doppler shift encountered by  a photon travelling from $U_*$ on $\scri^-$ to $u_*$ on $\scri^+$~\cite{details}.
Furthermore
\begin{equation}
U_H^* = U_* +
{1 \over \kappa_*}
\exp\left[  -\int_{-\infty}^{u_*} \kappa(\tilde u) \; d\tilde u \right],
\end{equation}
and
\begin{equation}
A_* = 
{1 \over \kappa_*}
\exp\left[  -\int_{-\infty}^{u_*} \kappa(\tilde u) \; d\tilde u +\kappa_* \; u_*\right].
\end{equation}
So we have explicit formulae for the parameters appearing in the exponential approximation.  

%\enlargethispage{20pt}

\noindent\underline{\emph{Surface gravity}:}  
The quantity $\kappa(u)$ is at this stage of the argument not in any sense a ``surface gravity'' --- it is just a specific way of encoding the functional relationship $U = p(u)$. Can we relate it to a surface gravity? Especially since we have emphasized that there is no need for an actual event horizon ever to form in this formalism?  
To see the relation to ``surface gravity'', we first use the fact that the (independent) coordinates $\{W,u\}$ provide a double-null cover of the relevant parts of the  spacetime: 
\begin{equation}
d s^2 = - F(W,u) \;d W \, d u + r(W,u)^2 \;d\Omega^2_{d-1}.
\end{equation}
Since $W$ and $u$ have been constructed to be affine on $\scri^\mp$ respectively then (assuming for the sake of argument no event horizon ever forms)
\begin{equation}
 F(+\infty,u) = 1 \qquad\hbox{and}\qquad F(W,-\infty) = 1,
 \end{equation}
which makes the metric simple on $\scri$. 

To now introduce a suitable notion of surface gravity note that the generators of $\scri^-$ are null geodesics, affinely parameterized by $W$.
Indeed, let us [in $(W, u, \theta_1, \dots, \theta_{d-1})$ coordinates] define the null vector
\begin{equation}
k^a = (1,0,0,\cdots,0),
\end{equation}
pointing in the direction of increasing $W$. The 4-acceleration
\begin{equation}
k^a \nabla_a k^b 
= \Gamma^b{}_{WW} 
= {F_{,W}\over F} (1,0,\cdots,0),
\end{equation}
enables us to identify a new and logically distinct quantity
\begin{equation}
\kappa_\mathrm{bulk}(W,u) = {F_{,W}\over F}.
\end{equation}
where this is now a ``bulk'' quantity defined everywhere in the spacetime, and not a ``surface gravity'' as such. Instead,  $\kappa_\mathrm{bulk}$ is a  ``bulk inaffinity estimator'' which measures the extent to which $W$ fails to be an affine parameter along null geodesics  of increasing $W$. It is this ``bulk inaffinity estimator''  that is closely related to textbook notions of surface gravity, while it is the ``peeling'' notion of $\kappa(u)$  that we have seen is related to the Hawking flux (this observation extends to even more general ``analogue spacetime'' settings, see for example~\cite{Macher}). 

If a true future-eternal event horizon forms then there will be a region of $\scri^-$ (namely $W>U_H$) which has no natural ``lift'' to $\scri^+$. Instead, this region of $\scri^-$ ``lifts'' to ${\cal H}^+$, the future horizon.  On this region we have the textbook definition of surface gravity
\begin{equation}
\kappa_\mathrm{inaffinity}(W)  = \lim_{u\to+\infty}  {\kappa_\mathrm{bulk}(W,u)}.
\end{equation}
If we make the further assumption that, after the future-eternal  event horizon forms, the black hole settles down to an asymptotically static state, then we have a relation
\begin{equation}
\kappa_H = \lim_{W\to+\infty} \kappa_\mathrm{inaffinity}(W) =  \lim_{u\to+\infty} \kappa(u).
\end{equation}
However, without such an asymptotic assumption the two notions are distinct. Even in this particular case they at best only coincide at $i^+$, future timelike infinity, and are unrelated at other locations. It is only for static (or stationary) spacetimes that the two notions exactly coincide for all times. (See~\cite{details} for details.)

The physics here is intriguing --- to get a Planckian flux not only do we not ever need the future horizon ${\cal H}^+$ to form, but the Hawking temperature is not logically or physically connected to the surface gravity of the horizon ${\cal H}^+$. Instead the Hawking temperature is primarily related to $\kappa(u)$, that is, the ``peeling off'' properties of the null geodesics that actually do reach $\scri^+$.
%It is only when a collapse settles down to a static black hole spacetime (the end point of classical collapse in GR) that these quantities happen to be equal.

\clearpage

\noindent\underline{\emph{Evaporating black holes}:}
Now consider a Schwarzschild black hole. 
It is easy to estimate $\kappa \sim  {M_P^2/ M}$ and  
$\dot\kappa \sim {M_\mathrm{P}^6 / M^4}.$
Thus, on the one hand the exponential approximation is valid as long as
\begin{equation}
|u-u_*| \ll t_{\rm cross} \; ({M/ M_P}), 
\end{equation}
($t_{\rm cross} = 1/\kappa$ is the time that light would take to travel a distance $2M$),
which for macroscopic black holes implies a very long timescale indeed, 
while on the other hand
\begin{equation}
{\dot\kappa/ \kappa^2} \sim
{M_\mathrm{P}^2/ M^2}.
\end{equation}
So the Hawking process for Schwarzschild black holes
does satisfy the ``adiabaticity condition'' we have enunciated above,
\emph{at least as long as the black hole is heavier than a few Planck
masses}. Physically the ``adiabaticity constraint'' is
equivalent to the statement that a photon emitted near the peak of the
Planckian spectrum, with $\hbar \omega_\infty\approx k T_H$, that is
$\omega_\infty \sim \kappa$, should \emph{not} see a large fractional
change in the peak energy of the spectrum over one oscillation of the
electromagnetic field. (That is, the change in spacetime geometry is adiabatic as
seen by a photon near the peak of the Hawking spectrum.) 

%\bigskip

\noindent{\underline{\em Discussion}:} 
We have demonstrated that any collapsing compact object (regardless of whether or not any type of horizon ever forms) will, provided the exponential approximation and adiabatic condition hold, emit a slowly evolving Planckian flux of quanta. A key observation is  that  it is the peeling function $\kappa(u)$ that controls the salient features of the Hawking flux whether or not a horizon ever forms. Even if  a future horizon forms,  the peeling function $\kappa(u)$ need not be directly related to its surface gravity.  Note that we have carefully described the Hawking flux as Planckian rather than thermal. To claim thermality one has to explicitly assume the formation of an event horizon (behind which one can hide correlations).  

As a byproduct, our analysis provides a reasonably complete physical picture of standard black hole evaporation that applies over most of the history of the collapse and evaporation process  (a picture that is broadly speaking compatible with the extant literature~\cite{lindesay+sheldon, bergmann+roman, ashtekar-bojowald, disinformation, hayward, hawking-dublin, hawking-post-dublin}). 
As is only to be expected, the physical picture developed in this Letter explicitly fails during the last few Planck times of the evaporation process, when the region of validity of the exponential approximation shrinks to Planck size, and when the adiabatic condition is explicitly violated.  (This need not be the case for compact horizonless objects such as quasi-black holes~\cite{fate, small-dark, revisit, ashtekar-bojowald, disinformation, hayward}.)  Note that if a horizon forms and then completely evaporates, then at the end-point $p(u)$ has a discontinuity, and $\kappa(u)$ will diverge. 
This strongly suggests that in this situation some form of ``thunderbolt'' will be emitted~\cite{thunderbolt,thunderbolt2}.

%\bigskip

%\enlargethispage{25pt}

\null
\vskip-20pt
%------------------------------------------------------------------------------

%------------------------------------------------------------------------------
\end{document}